\documentclass[debug,overfull,published]{epl}

\title{Luttinger liquid phenomenology and angle resolved \\ photoemission  
for single layer  
$\chem{Bi_2Sr_{2-x}La_xCuO_{6+\delta}}$\\ high--temperature superconductor}
\shorttitle{Luttinger liquid phenomenology}
\author{K.~Byczuk\inst{1}, C.~Janowitz\inst{2}, R.~Manzke\inst{2}, J.~Spa\l ek\inst{3}, 
\and W.~W\'ojcik\inst{4}}
\institute{
  \inst{1} Institute of Theoretical Physics, Warsaw University, ul. Ho\.za 69,
00-681 Warszawa, Poland\\
  \inst{2} Humboldt Universit\"at, Institute f\"ur Physik, Newtonstr. 15, 12489 Berlin, Germany \\
\inst{3} Institute of Physics, Jagiellonian University, ul. Reymonta 4, 30-059
Krak\'ow, Poland\\
\inst{4}Institute of Physics, Cracow University of Technology,
ul. Podchor\c{a}\.{z}ych 1, 30-084 Krak\.{o}w, Poland
} 

\begin{document}

\maketitle

\begin{abstract}
Recently observed  splitting in angular resolved photoemission spectroscopy (ARPES) on 
$\chem{Bi_2Sr_{2-x}La_xCuO_{6+\delta}}$ high--temperature superconductor 
(Janowitz~C. {\it et al.}, {\it Europhys. Lett.}, {\bf 60} (2002) 615) is
interpreted  
within the  phenomenological  Luttinger--liquid  framework, in which  both 
 the non--Fermi liquid scaling exponent of the spectral function and 
the spin--charge separation are introduced.  
The anomalous Green function with adjustable parameters fits very well to the Fermi edge 
and the low--energy part of ARPES along the $\Gamma-M$ line in the Brillouin 
zone. 
In contrast to one--dimensional models with Luttinger--liquid behavior we find that both 
the anomalous scaling $\alpha$ and the parameter $\delta$ describing the spin--charge 
separation are momentum dependent. 
The higher--energy part of the 
spectra is not accounted for by this simple Luttinger--liquid form of the 
Green function.  
In this energy regime additional scattering processes are plausible to produce the experimentally 
observed wide incoherent 
 background, which  diminishes as the inverse of the energy.  
\end{abstract}

The understanding of physical properties of high--temperature superconductors and, 
in particular, of their   
normal  phase properties, is a notoriously difficult problem for more than 15 years. 
The absence of the Fermi--liquid {\em quasiparticles} in the normal state at the
optimal doping  or in the underdoped regime,  
that was concluded on the 
basis of thermal, photoemission, or transport experiments \cite{htc}, has given
rise to the search for 
a {\em non--Fermi liquid  theory} of  high--temperature superconductors \cite{anderson}. 
Despite of  many microscopic and phenomenological approaches, the problem of finding the theory 
for high--temperature superconductors has not yet been  resolved.

In  a microscopic approach to many--body systems one investigates suitable correlation 
functions, usually within the renormalized perturbation scheme \cite{mahan}. 
From these functions  physical properties of the system are determined and
compared with  
 experimental results. 
In  a macroscopic (phenomenological) approach
one starts with a certain expression for either the thermodynamic potential (e.g. from a Ginzburg-Landau 
functional \cite{landau})   
or  correlation function \cite{inoue95,byczuk02}, which have adjustable 
parameters that are fixed by fitting the results to experimental data. 
Only in a limited number of cases the macroscopic approaches are traced back to microscopic 
theories. 
Nevertheless, the power of the phenomenological approach lies
 in its ability to explain physical properties of 
real materials and, sometimes, allows one to guess some features of the 
albeit unknown microscopic approach.

In the macroscopic {\em Fermi liquid} theory the proper single-particle excitations are 
the {\em quasiparticles},   
which remain in one--to--one correspondence with the states of noninteracting electrons and 
turn out to be the exact single--particle 
excitations near the Fermi surface at zero or low temperatures.  
In the microscopic theory these quasiparticles occur  as single poles
near the real axis on the complex energy plane in the retarded   
one--particle 
Green function $G(\vect{k},\omega)$, where
 $\vect{k}$ is a momentum and $\omega$ is an energy. 
These quasiparticles are seen as sharp peaks in the spectral function 
$A(\vect{k},\omega)\equiv -(1/\pi){\rm{Im }}G( \vect{k},\omega )$ appearing on the top of 
 an incoherent  background.
The angular resolved photoemission spectroscopy (ARPES) 
data taken on one-- and two--dimensional metallic systems 
can be interpreted as a direct probe of the spectral function 
$A(\vect{k},\omega)$, multiplied by the Fermi--Dirac occupation function, and convoluted with 
the apparatus function. 
This experimental technique was used to show that  $\chem{TiTe_2}$  is 
 a two-dimensional Fermi liquid  \cite{claessen}. 

The canonical example of the 
non--Fermi liquid theory is provided by the exactly solvable Tomonaga--Luttinger 
model of interacting electrons  propagating in one spatial dimension.
The  one--particle Green function in that case does not have any poles representing the electron--like
quasiparticles. 
Instead, it has  {\em branch cuts} on the real energy axis and is 
characterized by a non--universal 
exponent $\alpha$,
which depends on the strength of the electron--electron interaction. 
Physically speaking, if a single electron is either injected into or emitted from 
this one--dimensional system, it 
gets fractionalized into the whole bunch of exact one--particle excitations of this system, 
 the {\em holons} and the {\em spinons}, carrying respectively the charge and the spin.
Such an electronic system is called  the {\em Luttinger liquid}. 
The absence of quasiparticles in ARPES and the signatures of the spin--charge separation were
reported in  few experiments performed on effectively one-dimensional metallic compounds 
\cite{lld1a,lld1b,lld1c}. 
Recently, the spin--charge separation was suggested \cite{janowitz02} 
in ARPES carried out 
on a two--dimensional high--temperature superconductor. 

In the present Letter we investigate the ARPES data 
for $\chem{Bi_2Sr_{2-x}La_xCuO_{6+\delta}}$ high--temperature superconductor \cite{janowitz02} 
within a {\em macroscopic}  framework.
Namely, we  fit  the proposed phenomenological  the Luttinger--liquid Green function with few 
adjustable parameters to these  data.  
We show that those data can be consistently interpreted within the Luttinger liquid approach,
 in which the spin--charge separation occurs. 
We  find  that the parameters appearing in the one--particle Green function
are momentum dependent, which is  in contrast to the one--dimensional Tomonaga--Luttinger model, where
these parameters are universal.  
From this fitting procedure we determine the dispersion of the holons and the spinons. 
However, a high--energy--tail scaling as $1/\omega$ represents probably an incoherent background, which 
cannot be accounted for within our phenomenological scheme. 

The  one--particle Green function for a  Luttinger--liquid state 
in higher space dimensions may be proposed in the following  general form
\begin{equation}
G(\vect{k},\omega)=g\frac{ \omega_c^{\mu+\nu-1}e^{i\phi}}
{(\omega-\epsilon^c_{\vect{k}})^{\mu}(\omega-\epsilon^s_{\vect{k}})^\nu} \; ,
\label{eq1}
\end{equation}
where $\epsilon^c_{\vect{k}}$ ($\epsilon^s_{\vect{k}}$) is the momentum ($\vect{k}$) dependent 
charge (spin) one--particle excitation dispersion relation, $\mu$ and $\nu$ are $\vect{k}$--dependent
anomalous scaling exponents in the Luttinger liquid theory, $\omega_c$ is the energy 
cutoff, $\phi$ is a phase to be fixed, and $g$ is a normalization coefficient \cite{byczuk98}. 
When $\epsilon^c_{\vect{k}}= \epsilon^s_{\vect{k}}$  
the Luttinger--liquid 
Green function with only anomalous scaling 
is recovered \cite{yin96}. 
If $\mu=\nu=1/2$ the Luttinger--liquid Green function with only spin--charge 
separation is obtained. 
In the limit when both $\epsilon^c_{\vect{k}} = \epsilon^s_{\vect{k}}$ and $\mu = \nu = 1/2$, the expression  
(\ref{eq1}) represents the Fermi liquid with no damping of the quasiparticle states. 
In general case, the exponents $\mu$ and $\nu$ and the phase $\phi$ are 
to be constrained by other fundamental assumptions.

As in the low--energy Fermi liquid theory with linearized dispersion relation,  
 we assume that the spectral function obtained from 
(\ref{eq1}) is particle--hole symmetric, i.e. $A(\vect{k},\omega)=A(-\vect{k},-\omega)$ \cite{yin96}. 
Then, the following physically distinct cases should be considered: {\em (i)}~$\mu=\nu=1/2-\alpha$ and
{\em (ii)}~$\nu=1/2$ and $\mu=1/2-\alpha$, where $0\leq \alpha\leq1/2$ in both cases. 
In the case {\em (i)} both the holon and the spinon singularities on the complex energy plane 
diverge with the same exponent, which in turn leads to 
the same height of the two peaks appearing in the spectral function. 
The normal-- and the superconducting--state properties of the Luttinger liquid 
corresponding to the case {\em (i)} were investigated in Ref.~\cite{byczuk98}. 
However, the recent ARPES experiment \cite{janowitz02} 
has shown  that the heights of the holon and the spinon peaks are 
significantly different, that is  consistent with the case {\em (ii)}, also studied in Ref.~\cite{rodriguez00}. 
Guided by these experimental results,   
we choose the Luttinger--liquid Green function corresponding to  the case {\em (ii)}. 
In addition to the particle--hole symmetry, we postulate  the
absence of relevant and marginal interactions in the spin sector on the microscopic level. 
In  result, the spinon dispersion relation is not renormalized, i.e. 
$\epsilon^s_{\vect{k}}=\epsilon_{\vect{k}}$, where $\epsilon_{\vect{k}}$ is the 
bare--band dispersion relation. 
The holon dispersion relation is renormalized by the interactions and we assume
that $\epsilon^c_{\vect{k}} \equiv \delta(\vect{k})\epsilon_{\vect{k}}$,
where $\delta(\vect{k})\geq 1$ is a $\vect{k}$--dependent parameter measuring the magnitude of 
the spin--charge separation. 
Note that in the exact one--dimensional theory the parameters 
$\alpha$ and $\delta$ would not be momentum dependent.  

Our  phenomenological one-particle Green function can be rewritten in the form 
\begin{equation}
G(\epsilon_{\vect{k}},\omega)=
\frac{\omega_c^{-\alpha}e^{i\phi}}{\sqrt{(\omega-\epsilon^c_{\vect{k}})(\omega-\epsilon^s_{\vect{k}})}}
\frac{g(\alpha,\delta)}{(\omega-\epsilon^c_{\vect{k}})^{-\alpha}} \; , 
\end{equation}
where now $\alpha=\alpha(\vect{k})$ is the anomalous $\vect{k}$--dependent 
{\em scaling exponent} in the Luttinger
liquid theory. 
The particle--hole symmetry  fixes the unknown phase factor at the value 
$\phi=-\alpha \pi /2$.
The spectral function has then the form:
\begin{eqnarray}
A(\epsilon,\omega)=\frac{g(\alpha,\delta)\omega_c^{-\alpha}}{\pi}
\left\{ 
\sin\left(\frac{\pi \alpha}{2}\right)
\left[
\frac{\Theta(\omega-\epsilon_s)\Theta(\omega-\epsilon_c)}
{(\omega-\epsilon_s)^{1/2}(\omega-\epsilon_c)^{1/2-\alpha}}+
\frac{\Theta(\epsilon_s-\omega)\Theta(\epsilon_c-\omega)}
{(\epsilon_s-\omega)^{1/2}(\epsilon_c-\omega)^{1/2-\alpha}}
\right]\right. + \nonumber \\
\left.
\cos\left(\frac{\pi \alpha}{2}\right)
\left[
\frac{\Theta(\omega-\epsilon_s)\Theta(\epsilon_c-\omega)}
{(\omega-\epsilon_s)^{1/2}(\epsilon_c-\omega)^{1/2-\alpha}}+
\frac{\Theta(\epsilon_s-\omega)\Theta(\omega-\epsilon_c)}
{(\epsilon_s-\omega)^{1/2}(\omega-\epsilon_c)^{1/2-\alpha}}
\right]
\right\},
\label{eq3}
\end{eqnarray}
where $\epsilon_s=\epsilon\equiv \epsilon_{\vect{k}}$, 
$\epsilon_c\equiv \delta(\vect{k})\epsilon$,
 and  $\Theta(x)$ is  the  Heaviside step function. 
We fix the cutoff at 
$\omega_c=1\;\un{eV}$  to set the energy scale. 
Obviously, the results for $A(\epsilon , \omega )$ should not be dependent on 
the value of $\omega_c$.

The ARPES intensity, within a constant transfer matrix--approximation,  
is given by the  convolution
\begin{equation}
I(\omega)=I_0 \int \upd\omega' A(\epsilon,\omega')f(\omega'/kT)R(\omega'-\omega),
\end{equation}
where $T$ is the temperature, $f(x)$ is the Fermi function,  
and $R(\omega)$ is an apparatus resolution function assumed in 
  the  Gaussian form 
$
R(\omega)=\exp(-\omega^2/2\sigma^2)/\sqrt{2\pi\sigma^2}.
$
The phenomenological parameters $\alpha(\vect{k})$ and  $\delta(\vect{k})$ are determined 
by fitting $I(\omega)$ to the 
 ARPES data for $\chem{Bi_2Sr_{2-x}La_xCuO_{6+\delta}}$ compound with $x=0.4$ 
 for different values of the ARPES angle $\vartheta$ corresponding to
 different $\vect{k}$ \cite{janowitz02}. 
Also, the value of $\epsilon=\epsilon_{\vect{k}}$ for  given $\vect{k}$ is kept as a fitting 
parameter and thereby  exempting us from additional assumptions on the 
hopping integral amplitudes within a
tight--binding model of the electronic structure. 
We take $T=35\;\un{K}$ and  $\sigma=0.016\; eV$ \cite{janowitz02}. 

The $\chem{Bi_2Sr_{2-x}La_xCuO_{6+\delta}}$ high--temperature superconductor
($T_c=29\un{K}$ at the optimal 
doping $x=0.4$) is a prototype of a system with a single $\chem{CuO_2}$ layer. 
The large separation of the $\chem{CuO_2}$ layers makes the system 
essentially two-dimensional in the normal state;   
no bilayer splitting of the band is expected. 
The occurrence of the two--peak structure in ARPES along the $\Gamma-M$ direction 
in the Brillouin zone \cite{janowitz02} is interpreted as the 
Luttinger--liquid behavior in this system, as  discussed below.  
The splitting cannot be attributed to the $1 \times 4.8$ superstructure of the $\chem{CuO_2}$ planes, as the
split bands obtained in this manner cannot be fitted to the experiment.

\begin{figure}
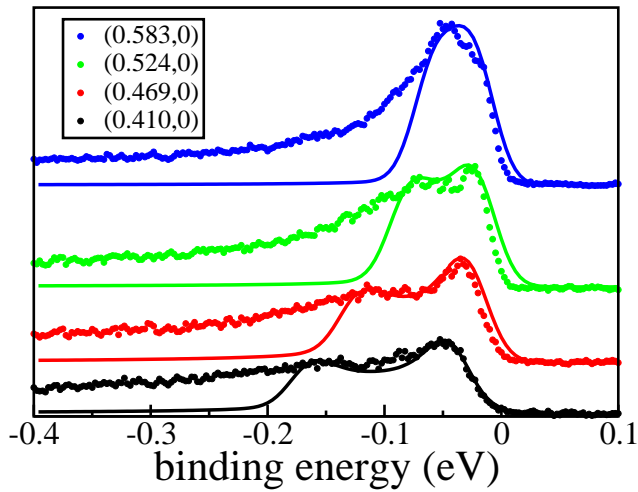

\vspace{0.50cm}
\onefigure[scale=0.36]{fit-spectra.eps}
\caption{Selected ARPES data  (dots) from Ref.~\cite{janowitz02} and the fitted spectral 
intensities (solid lines) 
within the Luttinger liquid phenomenology for different $|\vect{k}|$ vectors along 
$\Gamma -M$ direction of the Brillouin zone. The $|\vect{k}|$ vector values are in $\AA^{-1}$ units.}
\label{fig1}
\end{figure}

\begin{figure}
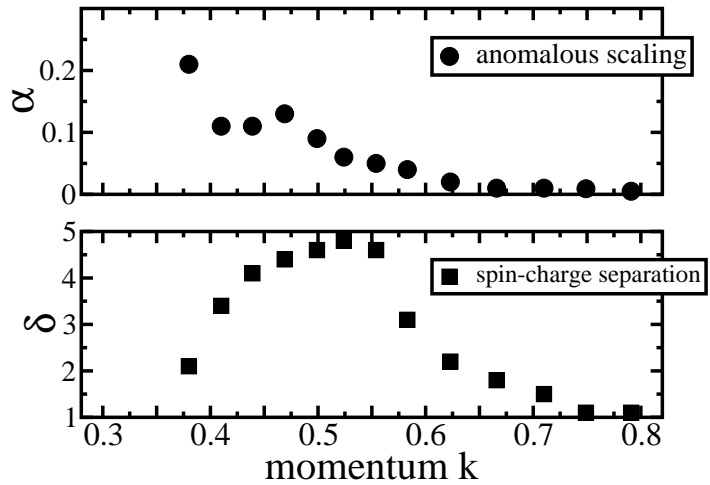

\vspace{0.50cm}
\onefigure[scale=0.36]{exponents.eps}
\caption{Anomalous scaling exponent $\alpha(\vect{k})$ (upper panel) and  
spin--charge separation parameter $\delta (\vect{k})$ (lower panel)
 as a function of momentum $|\vect{k}|$ (in $\AA^{-1}$) along the  $\Gamma -M$ direction.}
\label{fig2}
\end{figure}

Selected ARPES data along $\Gamma-M$ $[(0,0)-(\pi,0)]$ direction in the Brillouin zone \cite{janowitz02} and the 
fitted to them the intensity function $I(\omega )$,  are shown in Fig.~\ref{fig1}.
As one can see, the Fermi edge region and the two--peak structure are very 
well reproduced by the spectral function  form~(\ref{eq3}).
The higher energy tails are not accounted for by our low--energy Luttinger--liquid Green 
function \cite{comment}. 
In Fig.~\ref{fig2} we provide 
the $|\vect{k}|$ dependence of the phenomenological parameters $\alpha$ and $\delta$. 
The anomalous scaling exponent $\alpha$ is seen to decrease as one probes the system for 
higher $\vect{k}$ values along the $(0,0)-(\pi,0)$ direction.
The spin--charge separation parameter $\delta$ turns out to be non--monotonic
function of $\vect{k}$  
having a maximum around the middle point of the $\Gamma-M$ line.

\begin{figure}
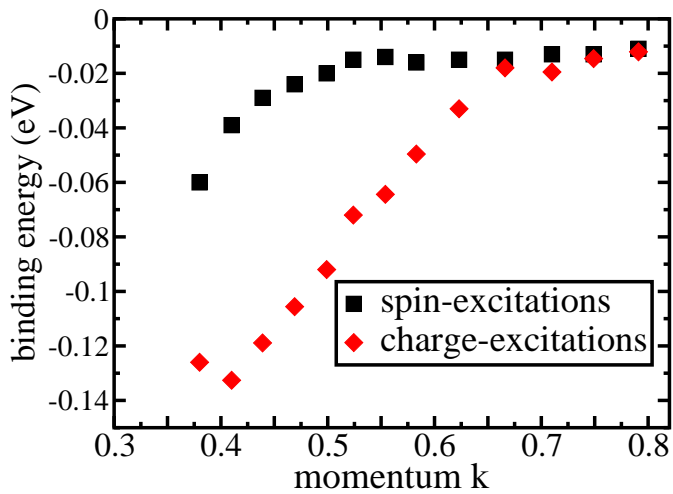

\vspace{0.50cm}
\onefigure[scale=0.36]{spin-charge-dispersion.eps}
\caption{Dispersion relations for spinon (squares) and holon (diamonds) excitations along the 
$\Gamma-M$ 
direction.}
\label{fig3}
\end{figure}

In Fig.~\ref{fig3} we plot the determined from the fitting 
dispersion relations $\epsilon^c_{\vect{k}}$ for holons and 
$\epsilon^s_{\vect{k}}$ for spinons along $\Gamma-M$ direction.  
Because of the interaction--induced renormalization, the holon excitations  
disperse more strongly then the spinons.
These two  {\em Luttinger--bands} are qualitatively 
similar to those obtained in \cite{janowitz02}, where they were represented by  
 two  separate Lorentzian spectral functions. 
A very small  difference is found near the $(\pi,0)$ point, where
$\epsilon^c_{\vect{k}}\approx  
\epsilon^s_{\vect{k}}$ in our case. 
This difference might be caused only  by a  numerical uncertainty of the two
procedures.

The presented comparison between the theory and the experiment 
 allows us  to conclude that  the ARPES data on 
$\chem{Bi_2Sr_{2-x}La_xCuO_{6+\delta}}$ compound are consistent with the
Luttinger liquid approach, in which  {\em both 
 the spin--charge separation $\delta$ and the anomalous scaling $\alpha$ are present 
at the same time}. 
This macroscopic investigation  might also cast some light onto a possible microscopic theory
of high-temperature superconductors.
Namely, in the one--dimensional systems of electrons interacting with short--range (screened) interaction,  
the Luttinger--liquid parameters $\alpha$ and $\delta$ were momentum independent. 
Here we find a significant momentum dependence of these parameters,
 that might suggest that the two--dimensional
theory of the correlated liquid would be more complex.  
This conjecture is also supported by the absence of the spin--charge splitting in the ARPES data along 
$\Gamma-Y$ direction in the same compound \cite{janowitz02}. 
In brief, one can say that the parameterization of the Green function  --  in terms of {\em branching cuts}  
(here) and in terms of {\em poles} (as in the Fermi liquid theory \cite{varma}) -- 
represent two {\em complementary} ways of viewing this two-dimensional system: 
the former represents the extrapolation of the concept of quantum liquid from one dimension, 
whereas the latter utilizes the extrapolation of the concept of the Fermi 
liquid from three dimensions.  
It is tempting to say that this two-sided possibility of viewing the electronic 
states in $\chem{CuO_2}$ planes means that the {\em lower critical dimensionality} 
for these system is $d_c=2$. 

Finally, we note that the phenomenological Green function (\ref{eq1}) seems to
describe only the low--energy part of the experimental spectra. 
The higher--energy spectral tails give almost one--half of the total spectral
weight, 
as might be inferred from the  fits in Fig.~\ref{fig1}. 
Moreover, we note that this higher energy tails diminish as $A(\epsilon,\omega)\sim 1/\omega$, which 
is slower then the expected asymptotic decay of the one-particle spectral function for interacting fermions. 
We do not believe that the $1/\omega$ dependence represents the asymptotic high--energy 
tail when $\omega \rightarrow \pm \infty$ \cite{mahan}. 
This is because the tail extends from $\omega \simeq 0.25 \; eV$, i.e. from the 
energy range, where the intrinsic many--body dynamics is still very important. 
It is plausible that other scattering mechanisms, not present in the strictly one--dimensional theory, 
are responsible for this slow decay of holons, as well as for  the broadening at higher
energies. 
This observation cannot be  quantified further within our phenomenological approach.
Nonetheless, it  might be used as  an  
additional constraint  in searching for the theory of high--temperature
superconductivity.  

In conclusion, we have presented the first quantitative analysis of photoemission spectra 
for the cuprates in terms of the Green function with branching cuts. 
The detailed analysis of the ARPES data for 
$\chem{Bi_2Sr_{2-x}La_xCuO_{6+\delta}}$ compound 
showed that they can be parameterized by a two--dimensional Luttinger liquid
approach. 
This approach   has  still to be modified with respect to its 
one--dimensional correspondent to account for the high--energy part of the
spectra. 
Also, the macroscopic parameters should be momentum dependent and additional scattering channels for
the charge excitations should be added to the present approach.

\acknowledgments

  Financial support of KB through the KBN Grant No. 
2P03B 082 24 and of JS and WW through the KBN Grant No. 2P03B 050 23
is  gratefully acknowledged. JS would like to thank the Foundation for Science
(FNP) for a senior fellowship for the years 2003-6.


\begin{thebibliography}{0}

\bibitem{htc}
For recent review see e.g. \Name{Ott H.R.}  in  
\Book{The Physics of Superconductors}
\Editor{ Bennemann K. H.  \and  Ketterson J. B.} 
\Vol{I} 
\Publ{Springer, Berlin} 
\Year{2003}
\Page{385ff}. 

\bibitem{anderson} 
\Name{Anderson P. W.} in 
\Book{Frontiers and Borderlines in Many-Body Physics, 
Proceedings of International School of Physics Varenna} 
\Publ{North Holland, Amsterdam} 
\Year{1987}; 
\Book{The Theory of Superconductivity in the High-$T_c$ Cuprates} 
\Publ{Princeton University Press, Princeton} 
\Year{1997}.



\bibitem{mahan}
  \Name{Mahan G.}
  \Book{Many--Body Physics}
  \Publ{Plenum Press, New York}
  \Year{1990}.

\bibitem{landau}
 See e.g. 
\Name{Blatter G. \and Geshkenbein V. B.} 
Ref. \cite{htc},  
\Page{p.725ff}.

\bibitem{inoue95}
\Name{Inoue I. H., Hase I., Aiura Y., Fujimori A., Haruyama Y., Maruyama T., \and Nishihara Y.}
\REVIEW{Phys. Rev. Lett.}{74}{1995}{2539}.

\bibitem{byczuk02}
\Name{Byczuk K., Bulla R., Claessen R., \and Vollhardt D.}
\REVIEW{Int. J. Mod. Phys. B}{16}{2002}{3759}.

\bibitem{claessen}
  \Name{Claessen R., Anderson R. O., 
Allen J. W., Olson C. G., Janowitz C., Ellis W. P., Harm S., Kalning M., Manzke R., \and Skibowski M.}
  \REVIEW{Phys. Rev. Lett.}{69}{1992}{808}.

\bibitem{lld1a}
\Name{Denlinger J. D., Gweon G. -H., Allen J. W., Olson C. G., Marcus J., Schlenker C., \and Hsu L. -S.}
\REVIEW{Phys. Rev. Lett.}{82}{1999}{2540}; 
\REVIEW{J. Phys. Chem. Solids}{56}{1995}{1849}; 
\REVIEW{Solid State Commun.}{123}{2002}{469}.  

\bibitem{lld1b}
\Name{Claessen R., Sing M., Schwingenschl\"ogl U., Blaha P., Dressel M., \and Jacobsen C.S.}
\REVIEW{Phys. Rev. Lett.}{88}{2002}{096402}.

\bibitem{lld1c}
\Name{Kim C., Matsuura A. Y., Shen Z. -X., Motoyama N., Eisaki H., 
Uchida S., Tohyama T., \and Maekawa S.} 
\REVIEW{Phys. Rev. Lett.}{77}{1996}{4054}. 

\bibitem{janowitz02}
  \Name{Janowitz C., M\"uller R., Dudy L., Krapf A., Manzke R., Ast C. R., \and  H\"ochst H.}
  \REVIEW{Europhys. Lett.}{60}{2002}{615}.

\bibitem{byczuk98}
  \Name{Byczuk K.,  Spa\l ek J., \and W\'ojcik W.}
  \REVIEW{Acta Phys. Polonica B}{29}{1998}{3871}.

\bibitem{yin96}
  \Name{Yin L. \and Chakravarty S.}
  \REVIEW{Int. J. Mod. Phys. B}{10}{1996}{805}.

\bibitem{rodriguez00}
  \Name{J. J. Rodriguez-Nunez, I. Tifrea, \and S. G. Magalhaes}
  \REVIEW{Phys. Rev B}{62}{2000}{4026}; 
\Name{Rodriguez-Nunez J.J., Budagodsky J.A., \and Tifrea M.I.} 
\REVIEW{Acta Phys. Polonica}{34}{2002}{383}.

\bibitem{comment}
Thus the spectral intensity was fitted only to a part of the experimental spectra 
to describe properly the low--energy features. In effect, higher energy tail was 
simply excluded in the fitting procedure.

\bibitem{varma}
\Name{Varma C. M., Littlewood P.B., Schmitt-Rink S., Abrahams E., \and Ruckenstein A.E.}
\REVIEW{Phys. Rev. Lett.}{63}{1989}{1996}.









\end{thebibliography}
\end{document}